\documentstyle[aps,epsfig,12pt]{revtex}
\tightenlines

\begin{document}
ADP-04-19/T600

JLAB-THY-04-25
\begin{center}
\vspace*{2cm} {\Large {\bf Liquid-gas phase transition in
nuclear matter including strangeness}}\\[0pt]
\vspace*{1cm} P. Wang$^a$, D. B. Leinweber$^a$, A. W. Thomas$^{a,b}$
and A. G. Williams$^a$ \\[0pt]
\vspace*{0.2cm} {\it $^a$Special Research Center for the Subatomic
Structure of
Matter (CSSM) and Department of Physics, University of Adelaide 5005,
Australia } \\[0pt]
\vspace*{0.2cm} {\it $^b$Jefferson Laboratory, 12000 Jefferson Ave.,
Newport News, VA 23606 USA}\\[0pt]
\end{center}

\begin{abstract}
We apply the chiral $SU(3)$ quark mean field model to study the
properties of
strange hadronic matter at finite temperature. The liquid-gas
phase transition is studied as a function of the
strangeness fraction.
The pressure of the system cannot remain
constant during the phase transition, since there are two independent
conserved charges (baryon and strangeness number). In a range
of temperatures around 15 MeV (precise values depending on the model
used)
the equation of state exhibits multiple bifurcates.
The difference in the strangeness fraction $f_s$ between the liquid
and gas phases is small when they coexist.
The critical temperature of strange matter turns out to be a
non-trivial function of the strangeness fraction.

\end{abstract}

\bigskip

\leftline{PACS number(s): 21.65.+f; 12.39.-x; 11.30.Rd}
\bigskip
\leftline{{\bf Keywords: Liquid-gas Phase Transition, Strange Hadronic
Matter,}}

{\bf ~~~~~~~~~~ Chiral Symmetry, Quark Mean Field}

\section{Introduction}

The determination of the properties of hadronic matter at finite
temperature and density is a fundamental problem in nuclear
physics. In particular, the study of the liquid-gas phase transition
in medium energy heavy-ion collisions is of considerable interest.
Many intermediate-energy collision experiments have been performed
\cite{Suraud} to investigate the
unknown features of the highly excited (hot) nuclei formed in
such collisions \cite{Panagiotou,Chen}. Theoretically, much effort has
been devoted to studying the equation of state for nuclear matter
and to discussing the critical temperature, $T_c$.
Recently, Natowitz $et$ $al.$ obtained the
limiting temperature by using a number of different experimental
measurements \cite{Natowitz1}. From these observations the authors
extracted the critical temperature of infinite nuclear matter,
$T_c=16.6\pm 0.86$ MeV \cite{Natowitz2}. We can expect that
further experiments may eventually yield the limiting temperature of
hypernuclei and the critical temperature for infinite strange hadronic
matter. It is therefore interesting to study the liquid-gas phase
transition of strange hadronic matter theoretically.

Exploring systems with strangeness, especially with large
strangeness fraction, has attracted a lot of interest in recent years.
Such a system has many astrophysical and cosmological
implications and is indeed interesting by itself.
There are many theoretical discussions for both strange hadronic
matter \cite{Schaffner,Wanga} and strange quark matter
\cite{Farhi}-\cite{Wang0}. However, most discussions
are at zero temperature. The properties of strange hadronic matter at
finite
temperature have not been studied very much yet.
Unlike symmetric nuclear matter, for strange hadronic
matter, there are two conserved charges, baryon number and strangeness.
Glendenning \cite{Glendenning} first discussed the phase transition
with more than one conserved charge
in general and applied it to the possible transition to quark
matter in the core of neutron stars. M\"uller and Serot \cite{Muller}
discussed
asymmetric nuclear matter, which has two conserved charges (baryon
number and isospin), using the stability conditions on the
free energy, the conservation laws and the Gibbs criterion for the
liquid-gas phase transition. The liquid-gas phase transition
of asymmetric nuclear matter was also discussed in
effective chiral models in Refs. \cite{Wang1,Qian}.
It was found that the critical temperature decreases with increasing
asymmetry parameter $\alpha$.

For strange hadronic matter, the method is similar to that for
asymmetric nuclear matter. In both cases there are two conserved charges.
Recently, Yang $et$ $al.$ \cite{Yang} used the extended
Furnstahl-Serot-Tang (FST) model \cite{FST}
to discuss the liquid-gas phase transition of strange hadronic matter.
The original FST model satisfies the $SU(2)$ chiral symmetry
and was first applied to study nuclear matter \cite{Furnstahl}.
In the extended FST model \cite{Yang}, the author found a so called
critical pressure above which the liquid-gas phase transition
cannot exist. As a result no critical strangeness fraction was
obtained for a given temperature. This critical pressure exists
in finite nuclei together with a limiting temperature, $T_{lim}$.
However, in infinite hadronic matter, physically, there should be
no such critical pressure, since the pressure of the system can
be much higher than the critical pressure.
For strange matter with $f_s=2$, i.e. pure $\Xi$ matter, from
Fig. 1 of Ref. \cite{Yang} one sees that the pressure does not increase
monotonically with density. This means that at this temperature,
pure $\Xi$ matter can be in liquid-gas phase coexistence.
Therefore, the binodal $p-\mu$ diagram at temperature $T=10$ MeV
should terminate at $f_s=2$. We will reconsider this problem and
show how the critical strangeness fraction can be obtained.

To study the properties of hadronic matter, we need
phenomenological models since QCD cannot yet be used directly.
The symmetries of
QCD can be used to constrain the hadronic interactions
and models based on $SU(2)_{L}\times SU(2)_{R}$
symmetry and scale invariance have been proposed.
These effective models have been
widely used to investigate nuclear matter and finite nuclei,
both at zero and at finite temperature \cite{Furnstahl}-\cite{Zhang2}.
Papazoglou $et$ $al.$ extended the chiral effective models to
$SU(3)_{L}\times
SU(3)_{R}$, including the baryon octets\cite{Papazoglou1,Papazoglou2}.
As well as models based on hadronic degrees of freedom,
there are some other models based on quark degrees of freedom, such as
the
quark meson coupling model \cite{Guichon,Kazuo},
the cloudy bag model \cite{Thomas},
the NJL model \cite{Bentz} and the quark mean field model \cite{Toki},
$etc.$.
Recently, we proposed a chiral $SU(3)$ quark mean field model and
investigated the properties of
hadronic matter as well as quark matter \cite{Wang3}-\cite{Wang6}.
This model is quite successful in describing the properties of
nuclear matter \cite{Wang3}, strange matter \cite{Wang4,Wang5},
finite nuclei and hypernuclei \cite{Wang6} at zero temperature.
In this paper, we will apply the chiral $SU(3)$ quark mean field
model to finite temperature and study the liquid-gas phase transition
of strange hadronic matter.

The paper is organized as follows. The model is introduced in section
II. In
section III we use the model to investigate strange hadronic matter
at finite temperature. The numerical results are discussed in section
IV and
section V summarises our findings.

\section{The model}

Our considerations are based on the chiral $SU(3)$ quark mean field
model (for details see Refs.~\cite{Wang4,Wang6}), which contains
quarks and mesons as basic degrees of freedom. Quarks are confined
in baryons by an effective potential. The quark meson interaction
and meson self-interaction are based on $SU(3)$ chiral symmetry.
Through the mechanism of spontaneous chiral symmetry breaking,
the resulting constituent quarks and mesons (except for
the pseudoscalars) obtain masses. The introduction of an explicit
symmetry breaking term in the meson self-interaction generates
the masses of the pseudoscalar mesons which
satisfy partially conserved axial-vector current (PCAC) relations.
The explicit symmetry breaking term of the
quark meson interaction leads in turn to reasonable hyperon potentials
in hadronic matter. For completeness, we introduce the main concepts of
the model in this section.

In the chiral limit, the quark field $q$ can be
split into left and right-handed parts $q_{L}$ and $q_{R}$:
$q\,=\,q_{L}\,+\,q_{R}$. Under $SU(3)_{L}\times SU(3)_{R}$ they
transform as
\begin{equation}
q_{L}^{\prime }\,=\,L\,q_{L},~~~~~q_{R}^{\prime }\,=\,R\,q_{R}\,.
\end{equation}
The spin-0 mesons are written in the compact form
\begin{equation}
M(M^{+})=\Sigma \pm i\Pi =\frac{1}{\sqrt{2}}\sum_{a=0}^{8}\left( \sigma
^{a}\pm i\pi ^{a}\right) \lambda ^{a},
\end{equation}
where $\sigma ^{a}$ and $\pi ^{a}$ are the nonets of scalar and
pseudoscalar mesons, respectively, $\lambda ^{a}(a=1,...,8)$ are the
Gell-Mann matrices, and $\lambda ^{0}=\sqrt{\frac{2}{3}}\,I$.
The alternative plus and minus signs correspond to $M$ and $M^{+}$.
Under chiral $SU(3)$ transformations, $M$ and $M^{+}$ transform as
$M\rightarrow M^{\prime }=LMR^{+}$ and $M^{+}\rightarrow
M^{+^{\prime }}=RM^{+}L^{+}$. In a similar way, the spin-1 mesons
are introduced through:
\begin{equation}
l_{\mu }(r_{\mu })=\frac{1}{2}\left( V_{\mu }\pm A_{\mu }\right)
= \frac{1}{2\sqrt{2}}\sum_{a=0}^{8}\left( v_{\mu }^{a}\pm a_{\mu }^{a}
\right) \lambda^{a}
\end{equation}
with the transformation properties:
$l_{\mu }\rightarrow l_{\mu }^{\prime }=Ll_{\mu }L^{+}$,
$r_{\mu }\rightarrow r_{\mu }^{\prime }=Rr_{\mu }R^{+}$.
The matrices $\Sigma$, $\Pi$,
$V_{\mu }$ and $A_{\mu }$ can be written in
a form where the physical states are explicit. For the scalar and
vector
nonets, we have the expressions
\begin{eqnarray}
\Sigma&=&\frac1{\sqrt{2}}\sum_{a=0}^8\sigma^a\lambda^a=\left(
\begin{array}{lcr}
\frac1{\sqrt{2}}\left(\sigma+a_0^0\right) & a_0^+ & K^{*+} \\
a_0^- & \frac1{\sqrt{2}}\left(\sigma-a_0^0\right) & K^{*0} \\
K^{*-} & \bar{K}^{*0} & \zeta
\end{array}
\right),
\end{eqnarray}
\begin{eqnarray}
V_\mu&=&\frac1{\sqrt{2}}\sum_{a=0}^8 v_\mu^a\lambda^a=\left(
\begin{array}{lcr}
\frac1{\sqrt{2}}\left(\omega_\mu+\rho_\mu^0\right) & \rho_\mu^+ &
K_\mu^{*+}\\
\rho_\mu^- & \frac1{\sqrt{2}}\left(\omega_\mu-\rho_\mu^0\right) &
K_\mu^{*0}\\
K_\mu^{*-} & \bar{K}_\mu^{*0} & \phi_\mu
\end{array}
\right).
\end{eqnarray}
Pseudoscalar and pseudovector nonet mesons can be written in a similar
fashion.

The total effective Lagrangian has the form:
\begin{eqnarray}
{\cal L}_{{\rm eff}} \, = \, {\cal L}_{q0} \, + \, {\cal L}_{qM} \, +
\,
{\cal L}_{\Sigma\Sigma} \, + \, {\cal L}_{VV} \, + \, {\cal L}_{\chi
SB}\,
+ \, {\cal L}_{\Delta m_s} \, + \, {\cal L}_{h}, + \, {\cal L}_{c},
\end{eqnarray}
where ${\cal L}_{q0} =\bar q \, i\gamma^\mu \partial_\mu \, q$ is the
free part for massless quarks. The quark-meson interaction ${\cal
L}_{qM}$
can be written in a chiral $SU(3)$ invariant way as
\begin{eqnarray}
{\cal L}_{qM}=g_s\left(\bar{\Psi}_LM\Psi_R+\bar{\Psi}_RM^+\Psi_L\right)
-g_v\left(\bar{\Psi}_L\gamma^\mu l_\mu\Psi_L+\bar{\Psi}_R\gamma^\mu
r_\mu\Psi_R\right)~~~~~~~~~~~~~~~~~~~~~~~  \nonumber \\
=\frac{g_s}{\sqrt{2}}\bar{\Psi}\left(\sum_{a=0}^8\sigma_a\lambda_a
+i\sum_{a=0}^8\pi_a\lambda_a\gamma^5\right)\Psi -\frac{g_v}{2\sqrt{2}}
\bar{\Psi}\left(\sum_{a=0}^8\gamma^\mu v_\mu^a\lambda_a
-\sum_{a=0}^8\gamma^\mu\gamma^5 a_\mu^a\lambda_a\right)\Psi.
\end{eqnarray}
In the mean field approximation, the chiral-invariant scalar meson
${\cal L}_{\Sigma\Sigma}$ and vector meson ${\cal L}_{VV}$
self-interaction terms are written as~\cite{Wang4,Wang6}
\begin{eqnarray}
{\cal L}_{\Sigma\Sigma} &=& -\frac{1}{2} \, k_0\chi^2
\left(\sigma^2+\zeta^2\right)+k_1 \left(\sigma^2+\zeta^2\right)^2
+k_2\left(\frac{\sigma^4}2 +\zeta^4\right)+k_3\chi\sigma^2\zeta
\nonumber \\ \label{scalar}
&&-k_4\chi^4-\frac14\chi^4 {\rm ln}\frac{\chi^4}{\chi_0^4}
+\frac{\delta}
3\chi^4 {\rm ln}\frac{\sigma^2\zeta}{\sigma_0^2\zeta_0}, \\
{\cal L}_{VV}&=&\frac{1}{2} \, \frac{\chi^2}{\chi_0^2} \left(
m_\omega^2\omega^2+m_\rho^2\rho^2+m_\phi^2\phi^2\right)+g_4
\left(\omega^4+6\omega^2\rho^2+\rho^4+2\phi^4\right), \label{vector}
\end{eqnarray}
where $\delta = 6/33$; $\sigma_0$, $\zeta_0$ and $\chi_0$ are the
vacuum
expectation values of the corresponding mean fields $\sigma$, $\zeta$
and $\chi$. The Lagrangian ${\cal L}_{\chi SB}$ generates the
nonvanishing
masses of pseudoscalar mesons
\begin{equation}\label{ecsb}
{\cal L}_{\chi SB}=\frac{\chi^2}{\chi_0^2}\left[m_\pi^2F_\pi\sigma +
\left(
\sqrt{2} \, m_K^2F_K-\frac{m_\pi^2}{\sqrt{2}} F_\pi\right)\zeta\right],
\end{equation}
leading to a nonvanishing divergence of the axial currents which in
turn satisfy
the relevant PCAC relations for $\pi$ and $K$ mesons. Pseudoscalar and
scalar
mesons as well as the dilaton field $\chi$ obtain mass terms by
spontaneous breaking of chiral symmetry in the Lagrangian
(\ref{scalar}).
The masses of the $u$, $d$ and $s$ quarks are generated by the vacuum
expectation values of the two scalar mesons, $\sigma$ and $\zeta$.
To obtain the correct constituent mass of the strange quark, an
additional mass term has to be added:
\begin{eqnarray}
{\cal L}_{\Delta m_s} = - \Delta m_s \bar q S q \, ,
\end{eqnarray}
where $S \, = \, \frac{1}{3} \, \left(I - \lambda_8\sqrt{3}\right) =
{\rm diag}(0,0,1)$, is the strangeness quark matrix.
Through these mechanisms,
the quark constituent masses are finally given by
\begin{eqnarray}
m_u=m_d=-\frac{g_s}{\sqrt{2}}\sigma_0
\hspace*{.5cm} \mbox{and} \hspace*{.5cm}
m_s=-g_s \zeta_0 + \Delta m_s,
\end{eqnarray}
where $g_s$ and $\Delta m_s$ are chosen to yield the constituent
quark mass in vacuum -- in our case, $m_u=m_d=313$ MeV and $m_s=490$
MeV. In
order to obtain reasonable hyperon potentials in hadronic matter,
it has been found necessary to include an additional coupling between
strange quarks and the scalar mesons
$\sigma$ and $\zeta$ \cite{Wang4}. This term is expressed as
\begin{eqnarray}
{\cal L}_h \, = \, (h_1 \, \sigma \, + \, h_2 \, \zeta) \, \bar{s} s \,
.
\end{eqnarray}
In the quark mean field model, quarks are confined in baryons
by the Lagrangian ${\cal L}_c=-\bar{\Psi} \, \chi_c \, \Psi$ (with
$\chi_c$
given in Eq. (\ref{Dirac}), below).
The Dirac equation for the quark field $\Psi_{ij}$, under the
additional
influence of the meson mean fields, is given by
\begin{equation}
\left[-i\vec{\alpha}\cdot\vec{\nabla}+\chi_c(r)+\beta m_i^*\right]
\Psi_{ij}=e_i^*\Psi_{ij}, \label{Dirac}
\end{equation}
where $\vec{\alpha} = \gamma^0 \vec{\gamma}$\,, $\beta = \gamma^0$\,,
the subscripts $i$ and $j$ denote the quark $i$ ($i=u, d, s$)
in a baryon of type $j$ ($j=N, \Lambda, \Sigma, \Xi$) and
$\chi_c(r)$ is a confining potential -- i.e. a static potential
providing confinement of quarks by meson mean-field configurations.
The quark effective mass, $m_i^*$, and energy $e_i^*$ are defined as
\begin{equation}
m_i^*=-g_\sigma^i\sigma - g_\zeta^i\zeta+m_{i0}
\end{equation}
and
\begin{equation}
e_i^*=e_i-g_\omega^i\omega-g_\phi^i\phi \,,
\end{equation}
where $e_i$ is the energy of the quark under the influence of
the meson mean fields. Here $m_{i0} = 0$ for $i=u,d$ (nonstrange quark)
and $m_{i0} = \Delta m_s = 29$~MeV for $i=s$ (strange quark).
Using the solution of the Dirac
equation~(\ref{Dirac}) for the quark energy $e_i^*$
it has been common to define
the effective mass of the baryon $j$ through the ans\"atz:
\begin{eqnarray}
M_j^*=\sqrt{E_j^{*2}- <p_{j \, cm}^{*2}>} \label{square}\,,
\end{eqnarray}
where $E_j^*=\sum_in_{ij}e_i^*+E_{j \, spin}$ is the baryon energy and
$<p_{j \, cm}^{*2}>$ is the subtraction of the contribution
to the total energy associated with spurious center of mass
motion. In the expression for the baryon energy $n_{ij}$ is the number
of quarks with flavor $"i"$ in a baryon
with flavor $j$, with $j = N \, \{p, n\}\,,
\Sigma \, \{\Sigma^\pm, \Sigma^0\}\,, \Xi \,\{\Xi^0, \Xi^-\}\,,
\Lambda\,$  and $E_{j \, spin}$ is the correction
to the baryon energy which is determined from a fit to the data for
baryon masses.
There is an alternative way to remove the spurious c.\ m.\ motion and
determine the effective baryon masses. In Ref.~\cite{Guichon2},
the removal of the spurious c.\ m. \ motion for three quarks moving in
a confining, relativistic oscillator potential was studied in some
detail. It was found that when an external scalar potential was
applied, the effective mass obtained from the interaction Lagrangian
could be written as
\begin{eqnarray}
M_j^*=\sum_in_{ij}e_i^*-E_j^0 \label{linear}\,,
\end{eqnarray}
where $E_j^0$ was found to be only very weakly dependent on the
external field strength.
We therefore use Eq.~(\ref{linear}), with $E_j^0$ a
constant, independent of the density, which is adjusted to give a
best fit to the free baryon masses.

Using the square root ans\"atz for the effective baryon
mass, Eq.~(\ref{square}), the confining potential
$\chi_{c}$ is chosen as a combination of scalar
(S) and scalar-vector (SV) potentials as in Ref.~\cite{Wang4}:
\begin{eqnarray}
\chi_{c}(r)=\frac12 [\,\chi_{c}^{\rm S}(r)
                         + \chi_{c}^{\rm SV}(r)\,]
\end{eqnarray}
with
\begin{eqnarray}
\chi_{c}^{\rm S}(r)=\frac14 k_{c} \, r^2 \,,
\end{eqnarray}
and
\begin{eqnarray}
\chi_{c}^{\rm SV}(r)=\frac14 k_{c} \, r^2(1+\gamma^0) \,.
\end{eqnarray}
On the other hand, using the linear definition of effective baryon
mass, Eq.~(\ref{linear}), the confining potential
$\chi_{c}$ is chosen to be the purely scalar potential $\chi_{c}^{\rm
S}(r)$.
The coupling $k_{c}$ is taken as
$k_{c} = 1$ (GeV fm$^{-2})$, which yields baryon mean square charge
radii (in the absence
of a pion cloud \cite{Hackett}) around 0.6 fm.

The properties of infinite nuclear matter and finite nuclei were
calculated with these two treatments of effective baryon mass in
Ref.~\cite{Wangcssm}. As we have explained there,
the linear definition of effective
baryon mass has been derived using a systematic relativistic approach
\cite{Guichon2}, while to the best of our knowledge no equivalent
derivation exists for the square root case. For high baryon density,
the predictions of these two treatments are quite different. Many physical
quantities change discontinuously at some critical density in the case of
square root ans\"atz, while the linear definition of baryon mass
yields continuous behavior for high density nuclear matter.
Both treatments of the spurious c.\ m.\ motion fit the saturation properties
of nuclear matter and therefore, for
densities lower than the saturation density,
these two treatments give reasonably similar results. In this paper, we will
discuss the liquid-gas phase transition of strange hadronic matter with
both treatments. We prefer the linear form because it has been derived. The
square root case is reported here because it is widely used and in
fact produces similar results in some regions. However, where they differ we
believe that the linear form is the more reliable.

\section{strange hadronic matter at finite temperature}

Based on the previously defined interaction,
the Lagrangian density for strange hadronic matter is written as
\begin{eqnarray}
{\cal L}&=&\bar{\psi}_B(i\gamma^\mu\partial_\mu-M_B^*)\psi_B
+\frac12\partial_\mu\sigma\partial^\mu\sigma+\frac12
\partial_\mu\zeta\partial^\mu\zeta+\frac12\partial_\mu
\chi\partial^\mu\chi-\frac14F_{\mu\nu}F^{\mu\nu} -
\frac14S_{\mu\nu}S^{\mu\nu}\nonumber \\
&&-g_\omega^B\bar{\psi}_B\gamma_\mu\psi_B\omega^\mu -g_\phi^B\bar{\psi}
_B\gamma_\mu\psi_B\phi^\mu +{\cal L}_M, \label{bmeson}
\end{eqnarray}
where
\begin{equation}
F_{\mu\nu}=\partial_\mu\omega_\nu-\partial_\nu\omega_\mu
\hspace*{.5cm} \mbox{and} \hspace*{.5cm}
S_{\mu\nu}=\partial_\mu\phi_\nu-\partial_\nu\phi_\mu.
\end{equation}
The term ${\cal L}_M$ represents the interaction between mesons which
includes the scalar meson self-interaction ${\cal L}_{\Sigma\Sigma}$,
the vector meson self-interaction
${\cal L}_{VV}$ and the explicit chiral symmetry breaking term
${\cal L}_{\chi SB}$, all defined previously.
The Lagrangian includes the scalar mesons
$\sigma$, $\zeta$ and $\chi$, and the vector mesons $\omega$ and
$\phi$.
The interactions between quarks and scalar mesons result in the
effective
baryon masses $M_B^*$, where subscript $B$ labels the baryon
$B = N, \Lambda, \Sigma$ or $\Xi$. The interactions between quarks and
vector mesons generate the baryon-vector meson interaction.
The corresponding vector coupling constants $g_\omega^B$ and $g_\phi^B$
are
baryon dependent and satisfy the relevant $SU(3)$ relationships.
In fact, we find the following relations for the vector coupling
constants:
\begin{equation}
g_\omega^\Lambda=g_\omega^\Sigma=2g_\omega^\Xi=\frac23g_\omega^N
\hspace*{.5cm} \mbox{and} \hspace*{.5cm}
g_\phi^\Lambda=g_\phi^\Sigma=\frac12g_\phi^\Xi=\frac{\sqrt{2}}3g_\omega^N.
\end{equation}

At finite temperature and density, the thermodynamic potential for
strange
hadronic matter is defined as
\begin{eqnarray}
\Omega &=& - \sum_{j=N, \Lambda, \Sigma, \Xi }\frac{g_j k_{B}T}
{(2\pi)^3}\int_0^\infty d^3\overrightarrow{k}\biggl\{{\rm ln}
\left( 1+e^{-(E_j^{\ast}(k) - \nu _j)/k_{B}T}\right) \\
&+& {\rm ln}\left( 1+e^{-(E_j^{\ast }(k)+\nu_j)/k_{B}T}
\right) \biggr\} -{\cal L}_{M},  \nonumber
\end{eqnarray}
where $E_j^{\ast }(k)=\sqrt{M_j^{\ast 2}+\overrightarrow{k}^{2}}$
and $g_j$ is the degeneracy of baryon $j$ ($g_{N, \Xi}=2$,
$g_\Lambda=1$
and $g_\Sigma=3$). The quantity $\nu _j$
is related to the usual chemical potential $\mu _j$ by $\nu _j
=\mu _j-g_{\omega }^j\omega -g_{\phi }^j\phi$.
The energy per unit volume and the pressure of the system can be
derived as $\varepsilon =\Omega -\frac1T
\frac{\partial\Omega}{\partial T}+\nu _j\rho_j$ and $p=-\Omega $,
where $\rho_j$ is the baryon density.

The mean field equation for meson $\phi _{i}$ is obtained by the
formula $\partial \Omega/\partial \phi_i=0$. For example,
the equations for $\sigma$, $\zeta$ are deduced as:
\begin{eqnarray}\label{eq_sigma}
k_{0}\chi ^{2}\sigma
-4k_{1}\left( \sigma ^{2}+\zeta ^{2}\right) \sigma -2k_{2}\sigma
^{3}-2k_{3}\chi \sigma \zeta -\frac{2\delta }{3\sigma }\chi ^{4}
+\frac{\chi^{2}}{\chi _{0}^{2}}m_{\pi }^{2}F_{\pi }  \nonumber \\
-\left( \frac{\chi }{\chi _{0}}\right) ^{2}m_{\omega }\omega ^{2}\frac{
\partial m_{\omega }}{\partial \sigma }+\sum_{j=N,\Lambda ,\Sigma ,\Xi
}
\frac{\partial M_{j}^{\ast }}{\partial \sigma } <\bar{\psi _{j}}\psi
_{j}>=0,~~~~~~~~~~
\end{eqnarray}
\begin{eqnarray}\label{eq_zeta}
k_{0}\chi ^{2}\zeta -4k_{1}\left(\sigma ^{2}+\zeta ^{2}\right)
\zeta -4k_{2}\zeta ^{3}-k_{3}\chi \sigma ^{2} -
\frac{\delta }{3\zeta }\chi ^{4}+\frac{\chi ^{2}}{\chi _{0}^{2}}
\left( \sqrt{2}m_{k}^{2}F_{k}-\frac{1}{\sqrt{2}}m_{\pi }^{2}F_{\pi }
\right)\nonumber\\
-\left( \frac{\chi }{\chi _{0}}\right)^{2}m_{\phi }\phi ^{2}
\frac{\partial m_{\phi }}{\partial \zeta }+\sum_{j=\Lambda ,\Sigma ,\Xi
}
\frac{\partial M_{j}^{\ast }}{\partial\zeta }
<\bar{\psi _{j}}\psi_{j}>=0,~~~~~~~~~~~~~~
\end{eqnarray}
where
\begin{equation}
<\bar{\psi}_j\psi_j>=\frac{g_j}{\pi ^{2}}\int_{0}^{\infty}
dk \frac{k^{2}M_j^{\ast }}{E_j^*(k)}
\left[n_j(k)+\bar{n}_j(k)\right].
\end{equation}
In the above equation, $n_j(k)$ and $\bar{n}_j(k)$ are the baryon
and antibaryon distributions, respectively, expressed as
\begin{equation}
n_j(k)=\frac{1}{exp\left[\left(E_j^*(k)-\nu_j\right)/k_B T\right]+1}
\end{equation}
and
\begin{equation}
\bar{n}_j(k)=\frac{1}{exp\left[\left(E_j^*(k)+\nu_j\right)/k_B
T\right]+1}.
\end{equation}
The equations for vector mesons $\omega$ and $\phi$ are expressed
as
\begin{equation}
\frac{\chi^2}{\chi_0^2}m_\omega^2\omega+4g_4\omega^3
=\sum_{j=N,\Lambda,\Sigma,\Xi}g_\omega^j\rho_j,
\end{equation}
\begin{equation}
\frac{\chi^2}{\chi_0^2}m_\phi^2\phi+8g_4\phi^3
=\sum_{j=\Lambda,\Sigma,\Xi}g_\phi^j\rho_j,
\end{equation}
where $\rho_j$ is the density of baryons of type $j$,
expressed as
\begin{equation}
\rho_j=\frac{g_j}{\pi ^{2}}\int_{0}^{\infty}
dk k^2\left[n_j(k)-\bar{n}_j(k)\right].
\end{equation}

Let us now discuss the liquid-gas phase transition. For
strange hadronic matter, we follow the thermodynamic approach
of Refs. \cite{Glendenning} and \cite{Muller}. The system will be
stable against separation into two phases if the free energy of
a single phase is lower than the free energy in all two-phase
configurations. This requirement can be formulated as
\cite{Muller}
\begin{equation}
F(T,\rho)<(1-\lambda)F(T,\rho^\prime)+\lambda F(T,\rho^{\prime\prime}),
\end{equation}
where
\begin{equation}
\rho=(1-\lambda)\rho^\prime+\lambda\rho^{\prime\prime}, ~~~~~
0<\lambda<1,
\end{equation}
and $F$ is the Helmholtz free energy per unit volume. The two phases
are denoted by a prime and a double prime.
If the stability condition is violated, a system with two
phases is energetically favorable. The phase coexistence is
governed by the Gibbs conditions:
\begin{equation}\label{eq_Gibbsmu}
\mu_j^\prime(T,\rho^\prime)=\mu_j^{\prime\prime}(T,\rho^{\prime\prime})
~~~~ (j=N,\Lambda,\Sigma,\Xi),
\end{equation}
\begin{equation}\label{eq_Gibbsp}
p^\prime(T,\rho^\prime)=p^{\prime\prime}(T,\rho^{\prime\prime}),
\end{equation}
where the temperature is the same in the two phases.
The chemical potentials of the baryons satisfy the following
relationship:
\begin{equation}\label{eq_mumu}
\mu_\Lambda=\mu_\Sigma=(\mu_N+\mu_\Xi)/2.
\end{equation}
Therefore, there are only two independent chemical potentials for the
four
kinds of baryons. They are determined by the total baryon
density, $\rho_B$ and the strangeness fraction, $f_s$, which are
defined as
$\rho_B=(\rho_N+\rho_\Lambda+\rho_\Sigma+\rho_\Xi)$ and
$f_s=(\rho_\Lambda+\rho_\Sigma
+2\rho_\Xi)/\rho_B$.

\section{Numerical results and discussions}

The parameters in this model were determined by the meson masses in
vacuum and the properties of nuclear matter which were listed
in table I of Ref. \cite{Wangcssm}.
We now discuss the liquid-gas phase
transition of strange hadronic matter. In Fig.~1, we plot the
pressure of the system versus baryon density for various
strangeness fractions, $f_s$, at temperature $T=15$ MeV for the square
root ans\"atz of the effective baryon mass (Eq. (\ref{square})). For
nonstrange
hadronic matter, the $p-\rho_B$ isotherms exhibit the form of two
phase coexistence with an unphysical region. The nuclear matter
can be in a state of liquid-gas coexistence at this temperature. With
increasing $f_s$, the pressure will increase. At a particular value of
$f_s$, the pressure will increase monotonically with increasing
density. As we will see later, the the strangeness fraction is
different in
the liquid and gas phases. Therefore, the system can still be
in liquid-gas coexistence, even though the pressure increases
monotonically with density. The unphysical region appears again
in the range of $1.0<f_s<1.6$. It is obvious that the behavior of
the pressure of strange hadronic matter is not monotonic with $f_s$.
For the linear definition of effective baryon mass (Eq.
(\ref{linear})), the results are plotted in Fig.~2.
At small strangeness fraction, say $f_s<0.4$, there are
unphysical regions. In the range $0.4<f_s<1.75$, the pressure
increases monotonically with increasing density, while for $f_s>1.75$,
the
unphysical regions appear again.

\begin{center}
\begin{figure}[hbt]
\centering{\
\epsfig{figure=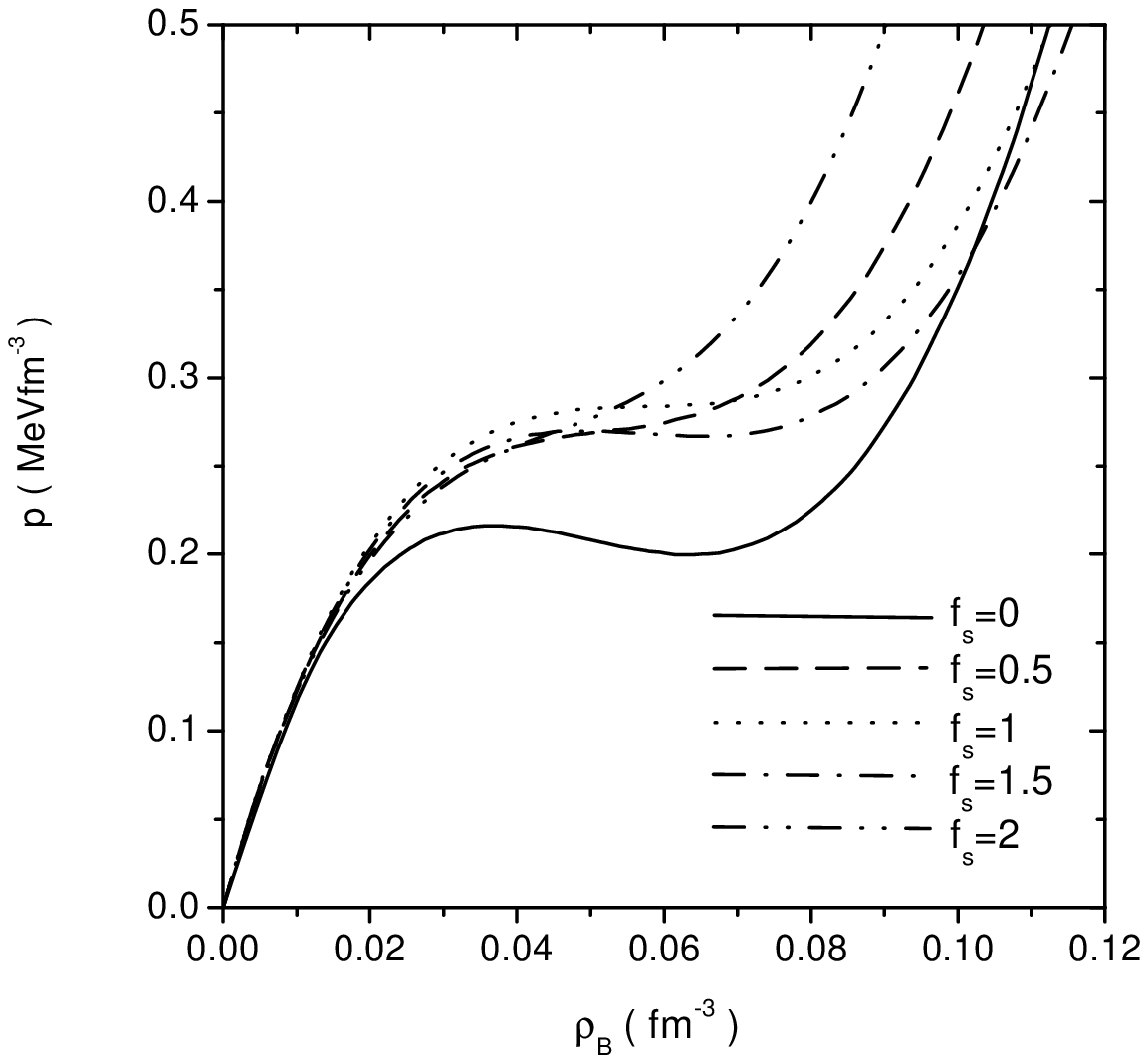,height=8cm}
}
\caption{The pressure of strange hadronic matter $p$ versus baryon
density $\rho_B$ with different strangeness fraction $f_s$
at temperature $T=15$ MeV in the case of square root ans\"atz
of effective baryon mass.}
\end{figure}
\end{center}

As we pointed out earlier, there are two independent chemical
potentials for the baryons. We now show how the Gibbs conditions
can be satisfied. As an example, we plot the chemical potentials
of nucleon and $\Lambda$ versus
$f_s$ at temperature $T=15$ MeV and pressure $p=0.23$ MeV-fm$^{-3}$
with the square root ans\"atz for the effective baryon mass
in Fig.~3 (For convenience, we use the reduced chemical potential
which is defined as $\widetilde{\mu}_j=\mu_j-M_j$).
The solid and dashed lines are for nucleon and $\Lambda$, respectively.
The Gibbs equations (\ref{eq_Gibbsmu}) and (\ref{eq_Gibbsp}) for phase
equilibrium demand equal pressure and chemical potentials for two
phases with different concentrations. The desired solution can be
found by means of the geometrical construction shown in Fig.~3, which
guarantees the same pressure and chemical potentials of nucleon
and $\Lambda$ in the two phases with different $f_s$.
Due to the chemical relationship between the baryons, the chemical
potentials of $\Sigma$ and $\Xi$ are also the same in the two phases.

\begin{center}
\begin{figure}[hbt]
\centering{\
\epsfig{figure=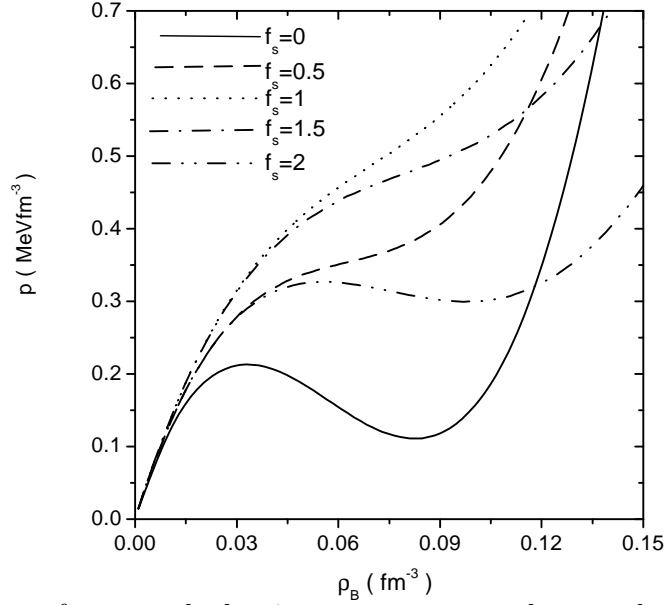,height=8cm}
}
\caption{The pressure of strange hadronic matter $p$ versus baryon
density $\rho_B$ with different strangeness fraction $f_s$
at temperature $T=15$ MeV in the case of linear definition
of effective baryon mass.}
\end{figure}
\end{center}

\begin{center}
\begin{figure}[hbt]
\centering{\
\epsfig{figure=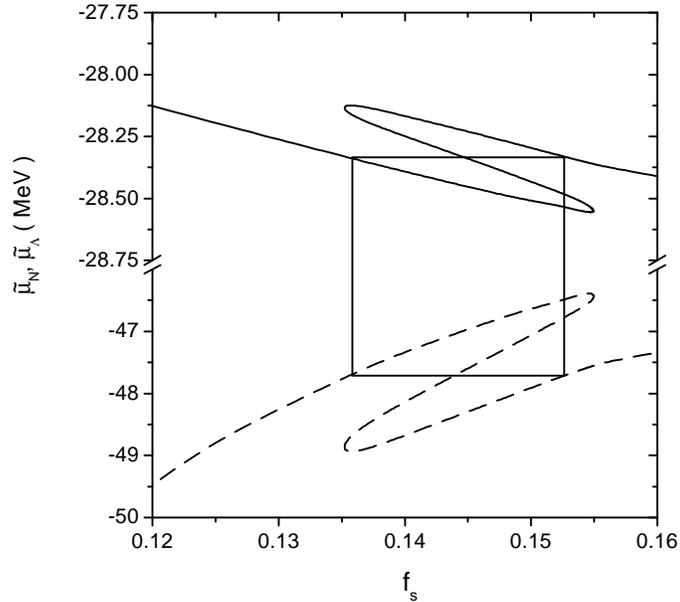,height=8cm}
}
\caption{Geometrical construction used to obtain the chemical
potentials and strangeness fraction in the two-phase coexistence
at temperature $T=15$ MeV and $p=0.23$ MeVfm$^{-3}$ The solid and
dashed lines are for nucleon and $\Lambda$, respectively.}
\end{figure}
\end{center}

For asymmetric nuclear matter, there is only one kind of
solution which satisfies Eqs. (\ref{eq_Gibbsmu}) and (\ref{eq_Gibbsp})
at the given pressure and temperature.
For hadronic matter, there is another solution of the Gibbs equations
with higher pressure at the same temperature. We show in Fig.~4 the
chemical potentials versus $f_s$ at pressure $p=0.27$ MeV-fm$^{-3}$ with
the same temperature as in Fig.~3. From the geometrical construction,
one can see that the difference of $f_s$ between the two phases
is very small. We can discuss in some detail the reason why
there is only one solution for asymmetric nuclear matter and
two solutions for strange hadronic matter. This is because for
asymmetric nuclear matter with $\alpha=(\rho_n-\rho_p)/(\rho_n+\rho_p)$,
the $\alpha$-dependence-behaviour of the pressure is
monotonic -- as can be seen clearly in Fig.~3 of Ref.~\cite{Wang1}.
However, in the case of strange hadronic matter, the dependence
on $f_s$ is not monotonic. For example, for temperature $T=15$
MeV, the point where $\partial p/\partial \rho_B =0$, appears in the
two regions of $f_s$ -- i.e. $0<f_s<0.2$ and $1.0<f_s<1.6$. This means
that the system can be in liquid-gas coexistence phase.

\begin{center}
\begin{figure}[hbt]
\centering{\
\epsfig{figure=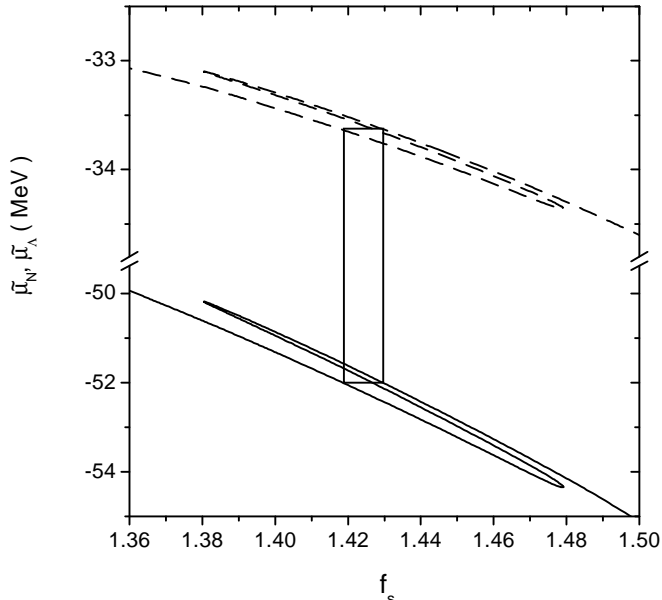,height=8cm}
}
\caption{Geometrical construction used to obtain the chemical
potentials and strangeness fraction in the two-phase coexistence
at temperature $T=15$ MeV and $p=0.27$ MeVfm$^{-3}$ The solid and
dashed lines are for nucleon and $\Lambda$, respectively.}
\end{figure}
\end{center}

The pairs of solutions shown in Fig.~3 with small strangeness fraction
form a binodal curve which is plotted in Fig.~5. There is a critical
point, A,
where the pressure is about 0.244 MeVfm$^{-3}$ with the corresponding
strangeness
fraction $f_s=0.24$.  The binodal curve is divided into two
branches by the critical point. One branch corresponds to the high
density (liquid) phase, the other corresponds to the low density
(gas) phase. Assume the system is initially prepared in the low
density (gas) phase with $f_s=0.2$. When the pressure increases to
some value, the two-phase region is encountered at point a and a
liquid phase at b with a low $f_s$ begins to emerge. As the system
is compressed, the gas phase evolves from point a to c, while the
liquid phase evolves from b to d. If the pressure of the system
continues to increase, the system will leave the
two-phase region at point d. The gas phase disappears and the system is
entirely in the liquid phase. This kind of phase transition is
different from the normal first order phase transition where the
pressure remains constant during phase transition.
If the strangeness is larger than 0.24, there is no phase
transition between liquid and gas phases. Therefore, for
a given temperature there exists a critical strangeness fraction,
above which the system can only be in the gas phase. In other words,
for a system with a fixed strangeness fraction $f_s$ there exists
a critical temperature, above which the system cannot change
completely into liquid phase however large the pressure.

\begin{center}
\begin{figure}[hbt]
\centering{\
\epsfig{figure=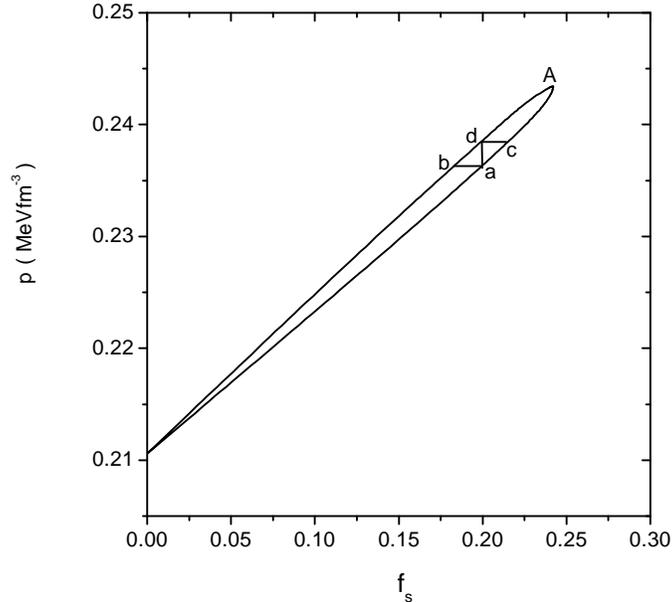,height=8cm}
}
\caption{The first binodal curve with smaller strangeness fraction
at temperature $T=15$ MeV. The points a through d denote
the liquid-gas phase transition. A is the critical point.}
\end{figure}
\end{center}

The solutions shown Fig.~4 with higher strangeness fraction
form another binodal curve and we plot it in Fig.~6.
As in Fig.~5, there are two branches divided by the critical
points B and C. One branch is for liquid phase , the other
is for gas phase . One can see that the difference of $f_s$
between liquid and gas phase is very small. At the critical points B
and C the strangeness fraction is about 1.04 and 1.66, respectively.
If $f_s$ is smaller than 1.04 but larger than 0.24, or $f_s>1.66$,
the system can only be in the gas phase.
At the critical points B and C, the strangeness fraction of
liquid and gas phase is the same. The liquid-gas phase
transitions at these two points are the same as the symmetric nuclear
matter. The pressure maintains the constant, $f_s$ does not
change during the phase transition.

\begin{center}
\begin{figure}[hbt]
\centering{\
\epsfig{figure=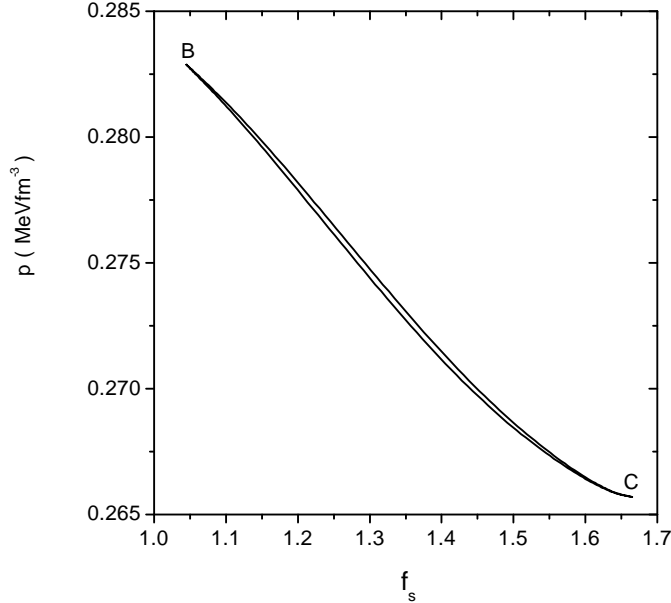,height=8cm}
}
\caption{The second binodal curve with larger strangeness fraction
at temperature $T=15$ MeV. B and C are the critical points.}
\end{figure}
\end{center}

For the linear case, there are also two kinds of solutions
which satisfy the Gibbs equations at $T=15$ MeV. However, the
third critical point C disappears because it moves to the point
with $f_s=2$. This is because at this temperature, the system
can be in the two phase coexistence when $f_s>1.75$. This
can be seen clearly from Fig.~2 where the unphysical region
exists for very large strangeness fraction $f_s$.

Therefore, for the case of square root ans\"atz, altogether
there are three critical strangeness fractions for strange
hadronic matter at temperature $T=15$ MeV, while for the linear
definition of the effective baryon mass, there are only two
critical points. In Fig. 7, we plot the critical temperature
versus strangeness fraction. For the square root case, the
critical temperature first decreases with increasing $f_s$.
When $0.65<f_s<1.3$, $T_C$ increases with $f_s$. For $f_s$ greater
than 1.3, $T_c$ decreases with $f_s$ again.
In the range, $14.6<T<15.3$ MeV,
there are three critical strangeness fractions for a given
temperature. For example, at $T=15$ MeV, the three critical values
of $f_s$ are about 0.24, 1.06 and 1.66. If $0.24<f_s<1.04$, the
strange hadronic matter can only be in the gas phase. When
$1.04<f_s<1.66$, the system can be in a state of liquid-gas
coexistence,
while for $f_s>1.66$, the system can once again only be in the gas
phase. The critical temperature for strange hadronic matter with
$f_s=2$ is about 13.6 MeV. For the case of the linear definition
of effective baryon mass, $T_c$ decreases with increasing
$f_s$ if $f_s$ is smaller than 1.1. If $f_s$ is larger than 1.1,
$T_c$ increases with $f_s$ till $f_s=2.0$. Compared with the
square root case, $T_c$ changes more quickly with $f_s$ in this case.
The
highest and lowest $T_c$ is 17.9 and 12.0 MeV with the
corresponding $f_s=0$ and $f_s=1.1$.

\begin{center}
\begin{figure}[hbt]
\centering{\
\epsfig{figure=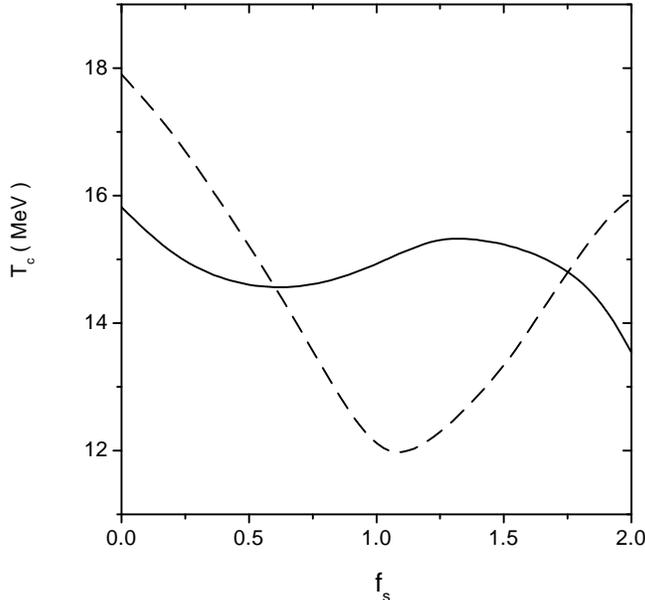,height=8cm}
}
\caption{The critical temperature $T_c$ versus strangeness fraction
$f_s$. The solid and dashed lines correspond to the square root
and linear case, respectively.}
\end{figure}
\end{center}

\section{Summary}

In this paper, we applied the chiral $SU(3)$ quark mean field model
to investigate the properties of strange hadronic matter at finite
temperature and density. All the parameters have been determined
in earlier papers and there is no further parameter to be adjusted.
The model works very well at zero temperature. The saturation
properties and compression modulus of nuclear matter are reasonable.
The
hyperon potentials are close to the empirical values for strange
hadronic matter. The results of finite nuclei and hypernuclei are
also consistent with the experiments. In this paper, the liquid-gas
phase
transition of strange hadronic matter was
studied in this model. For strange hadronic matter, there are two
independent conserved charges. The system will be in the liquid-gas
phase coexistence if the pressure and the chemical potentials of all
the
baryons are the same in the two phases. During the phase transition,
the strangeness
fraction $f_s$ of liquid and gas phases is different and
changes during phase transition, though the total $f_s$ is
conserved. We found that there are two branches of
solutions which satisfy the Gibbs equations at some range of
temperature.
One corresponds the phase transition at small $f_s$, while
the other corresponds the phase transition at large $f_s$.
For the square root ans\"atz of effective baryon mass,
there are three critical strangeness fractions during
$14.6<T<15.3$ MeV.
For the linear definition of effective mass, there
are two critical points when $12.0<T<15.9$ MeV
since the third one disappears and moves
to the point with $f_s=2$.
The difference of $f_s$ in the two phases is small,
especially in the case of higher strangeness fraction case.
The critical temperature $T_c$ does not change monotonically with
$f_s$.
For the square root case, if $f_s<0.65$ or $f_s>1.3$, $T_c$ decreases,
while for $0.65<f_s<1.3$, $T_c$ increases with
increasing $f_s$. For the linear case, $T_c$ first decreases and then
increases
with increasing $f_s$. The minimum critical temperature is about
12.0 MeV with $f_s=1.1$.

\bigskip
\bigskip

\section*{Acknowledgements}
This work was supported by the Australian Research Council
and by DOE contract DE-AC05-84ER40150, under which SURA operates
Jefferson Laboratory.

\end{document}